\documentclass[10pt, a4paper, conference, compsocconf]{IEEEtran}
\ifCLASSINFOpdf
   \usepackage[pdftex]{graphicx}
\else
\fi
\usepackage{amssymb}
\usepackage{amsthm}
\usepackage{graphicx}

\newtheorem{definition}{Definition}
\begin{document}
%
\title{Does diversity of papers affect their citations?  Evidence from American Physical Society Journals }


\author{\IEEEauthorblockN{
Murali Krishna Enduri\IEEEauthorrefmark{1},
I. Vinod Reddy\IEEEauthorrefmark{1}and
Shivakumar Jolad\IEEEauthorrefmark{1}  }
\IEEEauthorblockA{\IEEEauthorrefmark{1}Indian Institute of Technology Gandhinagar\\
VGEC Campus, Chandkheda\\
Ahmedabad, Gujarat 382424, INDIA
\\ Email:   \{endurimuralikrishna, reddy\_vinod, shiva.jolad\}@iitgn.ac.in }
}


%


\maketitle

\begin{abstract}
In this work, we study the correlation between interdisciplinarity of papers within physical sciences and their citations by using meta data of articles 
published in  American Physical Society's Physical Review journals between 1985 to 2012. We use the Weitzman diversity
index to measure the diversity of papers and authors, exploiting the hierarchical structure of PACS (Physics and Astronomy Classification Scheme) codes. 
We find that the fraction of authors with high diversity is increasing with time, where as the fraction  of least diversity are decreasing, and moderate diversity 
authors have higher tendency to switch over to other diversity groups. The diversity index of papers is correlated with the citations they received in a 
given time period from their publication year. Papers with lower and higher end of diversity index receive lesser citations than the moderate diversity papers.  
\end{abstract}


%
\IEEEpeerreviewmaketitle

\section{Introduction}
In the second  half of the twentieth century many scientific disciplines became more interconnected. Scientists working in varied fields felt the need 
to connect with people across different disciplines to study phenomena which required insights and expertise from multiple fields \cite{stehr2000practising}. This process has 
been accelerated in the last three decades as the advent of internet made people instantly discover the work done by other people world wide, and interact 
with people irrespective of their geographic location. Digital revolution has also enabled generation of enormous data on scientific articles published by
various journals. This generated tremendous interest among scientists working in scientometric and bibliometric studies, to understand the evolution of scientific disciplines, measure the impact of  articles through citation analysis, 
discover the pattern of scientific  collaboration etc \cite{bordons2005analysis, zitt2005facing, van2002assessment}. In recent years, many studies have 
focused on understanding  the  interdisciplinarity through variety of approaches \cite{porter2006interdisciplinary},  \cite{huutoniemi2010analyzing}, \cite{wagner2011approaches}, 
\cite{leydesdorff2011indicators}, and \cite{shi2011diversity}, but primarily using citation data and measures based on entropy, simpson index etc \cite{rafols2010diversity}. 
A recent special issue of \emph{Nature} summarizes the work on measuring interdiscplinarity and tracks the trend 
in interdisciplinary work across different fields from 1950 onwards \cite{Richard2015inter}. 
General conclusion has been that the interdisciplinary research is on the rise, especially since mid 1980's, they take time to have an impact and too 
much interdisciplinarity can decease the citations received.   

Interdisciplinary work is characterized by the diversity of inputs from different fields that contribute to making it. 
Consequently measuring the diversity of paper (author) captures the degree of interdisciplinarity of work (person). Diversity 
measures can then be used to understand whether more diverse papers generate greater impact? , are authors who work in diverse 
research areas necessarily have better publication record with higher citations? Thanks to the availability of large data sets of 
journal papers, such questions can be addressed. For example: using DBLP database of computer science, Chakraborty \emph{et al.}  \cite{chakraborty2015understanding} 
studied the diversity of 
researcher's scientific publications to understand the features that lead to triumphant career, and using co-citation cluster analysis on electrochemistry journal database Schmidt \emph{et al.} \cite{schmidt2006methodological} studied dynamics of diversity across six different countries in electrochemistry. Diversity measures can be used to characterize 
behaviors of individuals in complex networks. Lu Liu \emph{et al.} \cite{liu2010mining} gave an efficient algorithm to find the top-$k$ diverse 
nodes on a dynamic network.  Quan Shi \emph{et al.} \cite{shi2011diversity} found that local (global) diversity of authors in DBLP network tend 
to decay as an exponential (Gaussian) distribution. They also found that authors with more diverse social ties are more competitive. 
The interdisciplinarity in physical science research has been studied by Pan \emph{ et al.} using American Physical Society journals (APS) 
\cite{pan2012evolution}, and  observed that over time from 1980's there is a steady increase in interactions between the different fields 
and subfields of Physics. 
Chakraborty \emph{et al.} \cite{chakraborty2013automatic} developed supervised classification model to distinguish between core and 
interdisciplinary fields in DBLP database and studied their evolution and impact on the field. Using the APS Physical review database, 
Martin \emph{et al.} \cite{martin2013coauthorship} and Redner \emph{et al.} \cite{redner2005citation} studied correlation between authorship 
and citation, and found that individuals cite their collaborators work more quickly  compared to others work.

In this work, we study the interdisciplinarity of papers and authors using  Weitzman diversity \cite{page2010diversity} 
measure on hierarchical structure of Physics and Astronomy Classification Scheme (PACS) codes of papers published in American 
Physical Society (APS) journals. We examine the relation between the diversity and citation of papers and authors. We discovered 
that papers with extreme low and high diversity receive low citations compared to papers with moderate diversity.

\section{Dataset}
\label{sec:Dataset}
The American Physical Society (APS) started publishing Physical Review journal from 1893. APS added other journals like Reviews of
Modern Physics (1929), Physical Review Letters (1958), Physical Review A,B,C and D (1970), Physical Review E (1993) and most recently 
Physical Review X in 2011. 
In this paper we use all scientific papers published in APS 
Physical Review (PR) Journals (Physical Reviews A through E, Review of Modern Physics and Physical Review letters) from 1985 to 2012 
to study diversity profile and citations. For each paper, the data set contains unique digital object identifier (DOI), paper title, authors 
of paper, date of publication, affiliations of each author, Physical Review references of the paper and PACS codes. Along with the meta data of 
journal papers, we also have citations of papers published in APS from journals published in Physical Review (hence excludes citations received 
from non APS journals).  In Table.\ref{tab:BasicStats}, we show the basic descriptive statistics of the data.

\begin{table}[ht!]
\centering
\caption{Basic Statistics of Data 1985-2012}
\label{tab:BasicStats}
\begin{tabular}{| l | l | l}
\hline
Number of authors & 343055     \\
\hline
Number of papers  & 399713     \\
\hline
Average number of papers by an author & 9.07  \\
\hline
Average number of authors per paper  & 7.59      \\
\hline
Average number of PACS codes per author  & 10.04      \\
\hline
Average number of PACS codes per paper  & 2.92      \\
\hline
Average diversity of author   & 13.16  \\
\hline
Average diversity of paper  & 3.59  \\
\hline
Average citation per paper  & 10.22  \\
\hline
\end{tabular}
\end{table}

\begin{figure}[ht!]
\centering
  \includegraphics[trim=2cm 0.6cm 3.3cm 2.2cm, clip=true, scale=0.35]{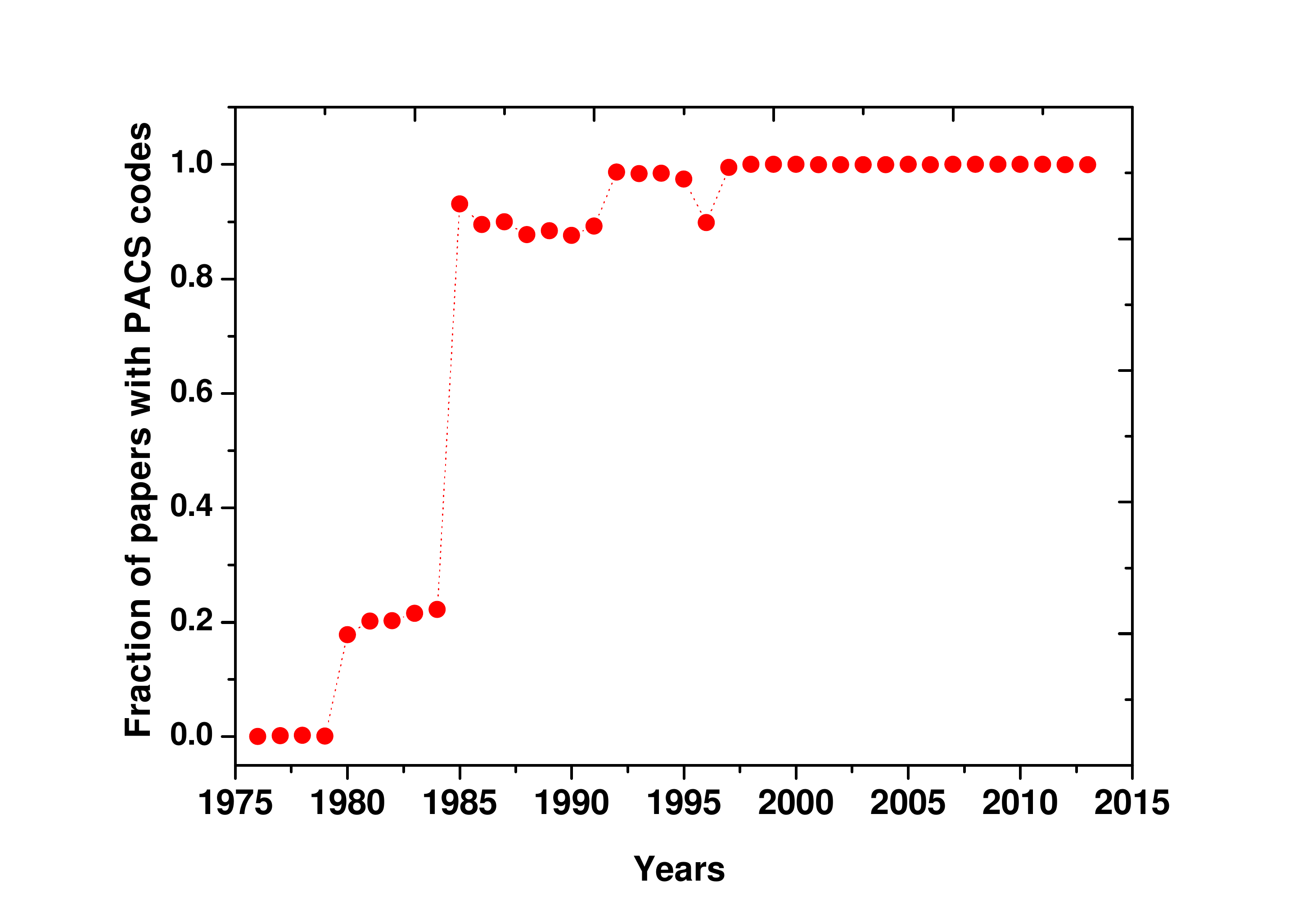}
\caption{Fraction of papers with PACS codes from 1975 to 2012}
\label{fig:FractionofpaperwithPACS} 
\end{figure}

\subsection{PACS classification}
PACS is a hierarchical classification scheme representing  different field and subfields of Physics up to five levels. A PACS 
code consists of two pairs of numbers followed by a pair of non numeric characters, separated by dots. 
For example in PACS code $04.25.dg$, the first digit 0 represents {\it General} Physics, 4 - {\it General relativity and gravitation}, 
25 - {\it Approximation methods; equations of motion } and $d$ represents {\it Numerical relativity} and $g$ represents  
{\it Numerical studies of black holes and black-hole binaries}. 
PACS codes are regularly revised and updated overtime by American Institute of Physics (AIP),  new codes are 
introduced and some codes are deleted. In our analysis we consider PACS codes up to third level (first four digits) of hierarchy as 
they are reasonably stable upto this level and represents all subfields of physics. We ignore the higher level hierarchy to maintain consistency PACS codes of all papers in our analysis.  PACS codes were introduced 1975 and in use since then. But large fraction of papers published between 1975 and 1984 have not assigned any PACS codes (see Fig. \ref{fig:FractionofpaperwithPACS}). 
We choose the period from 1985 onwards, as the compliance towards PACS code jumped to more than 90\% and have been consistently high since then. 
 
\begin{figure}[ht!]
  \centering
  \includegraphics[trim=2cm 0.5cm 3.3cm 2.2cm, clip=true, scale=0.35]{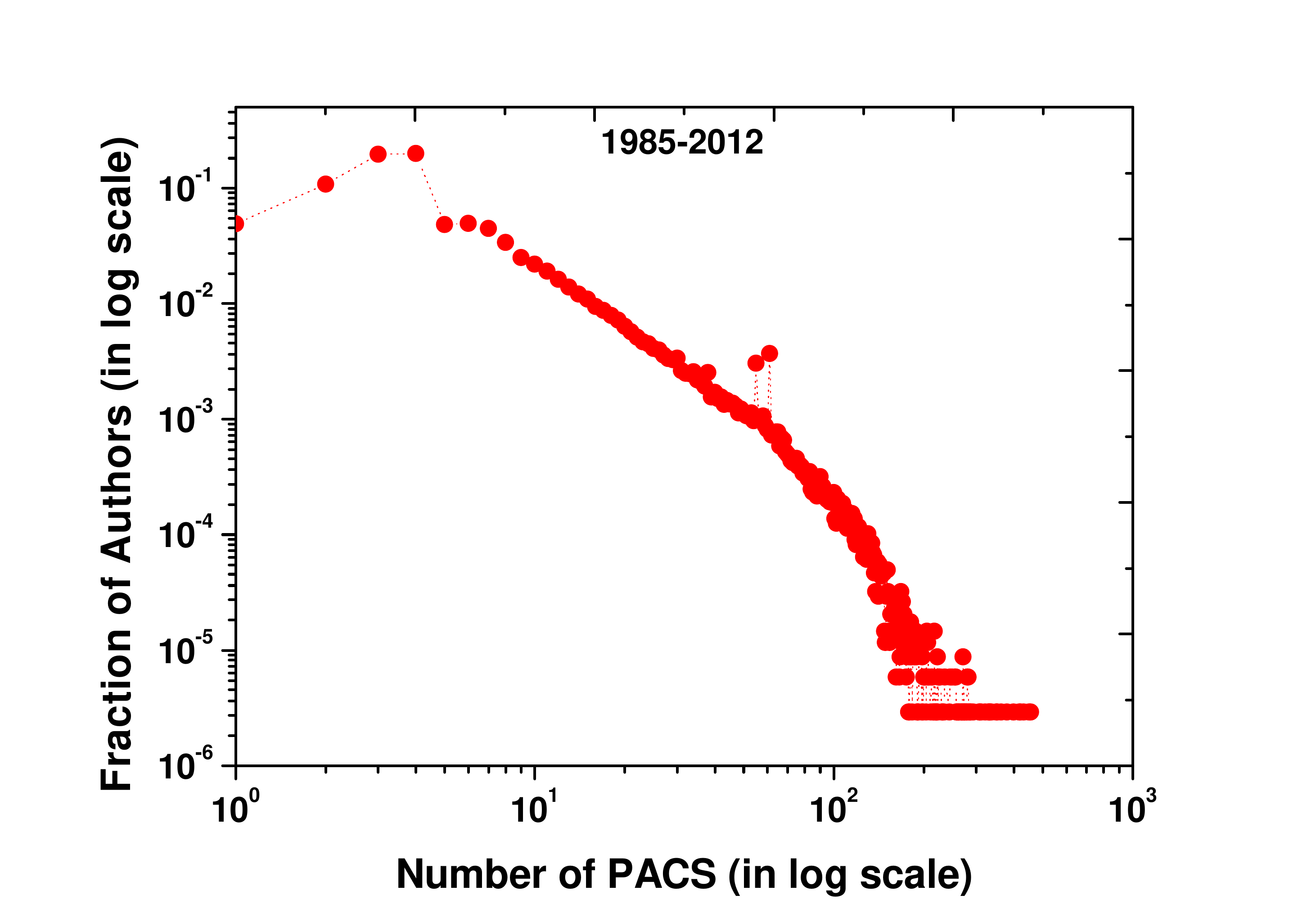}
\caption{Fraction of authors contributing to different number of PACS codes.}
\label{fig:Fraction of Authors} 
\end{figure}

\begin{figure}[ht!]
  \centering
  \includegraphics[trim=2cm 0.6cm 3.3cm 1.9cm, clip=true, scale=0.35]{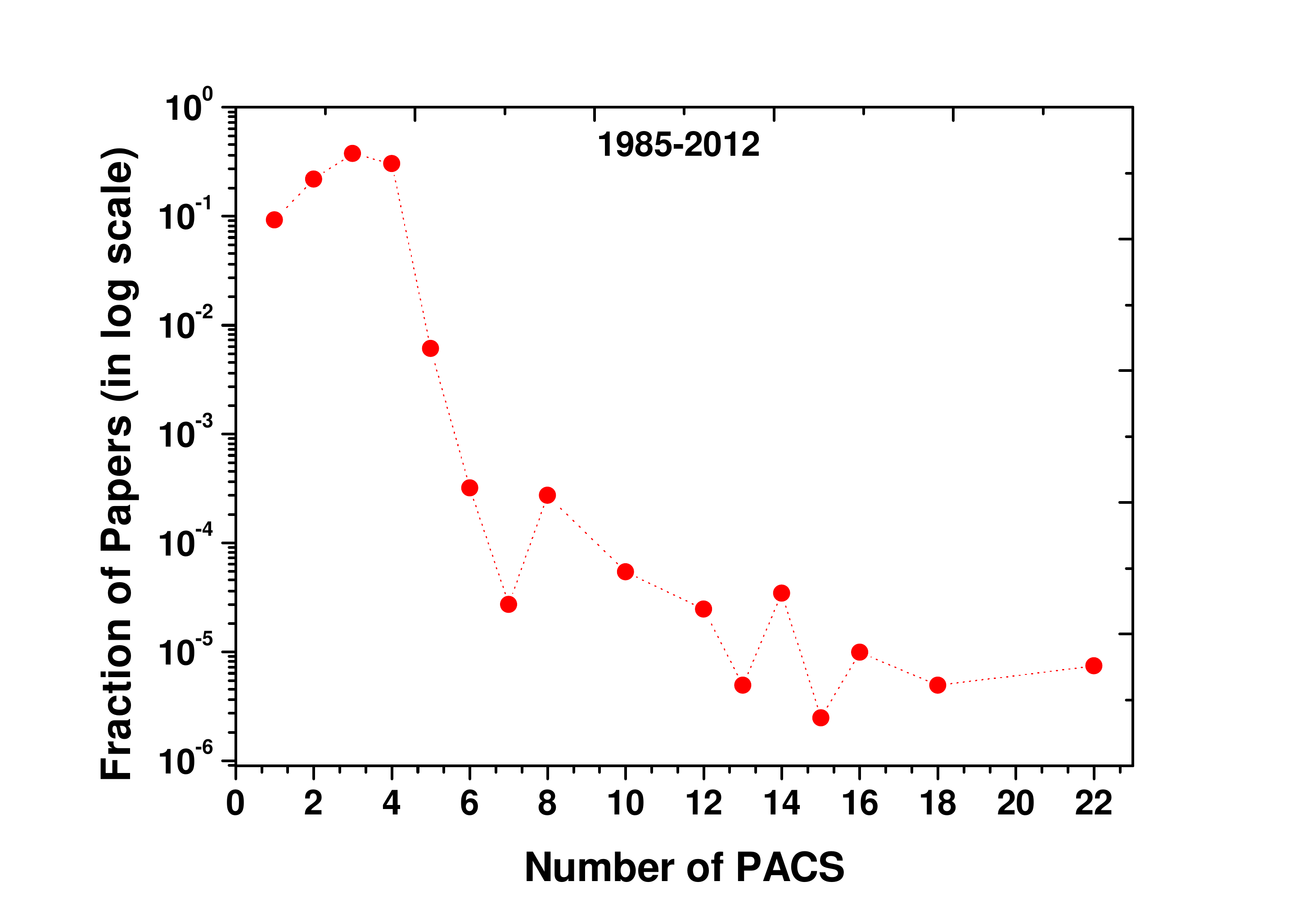}
\caption{Fraction of papers contributing to different number of PACS codes.}
\label{fig:Fraction of papers} 
\end{figure}

In Fig. \ref{fig:Fraction of Authors}, we show the fraction of authors using different number of 
PACS codes (plotted in log-log scale) in the papers published between 1985-2012. Large fraction of authors have used only 1 to 4 PACS codes. 
 We can observe a power law decay till PACS of 60, there after, we see the slope changing. The overall pattern seems  
to follow double pareto distrbution \cite{reed2004double}, but detailed study is yet to be carried out. Distribution of papers using different 
PACS codes  (in Fig.\ref{fig:Fraction of papers}) does not show any specific trend and reaches peak at four PACS codes, 
local minima at seven and there after it is fluctuating. 

\section{Diversity}

PACS data contains rich information on the multiple fields and subfields a papers addresses. 
The hierarchical structure of PACS can be exploited to understand the diversity of papers and authors.
Various measures based Shannon entropy, Simpson index, Gini-Simpson index  have been used to study the diversity in 
bibliographic studies \cite{rafols2010diversity} \cite{schmidt2006methodological}.  These studies primarily use citations of  papers 
received from journals of different fields as inputs to these measures. They implicitly assume that diversity 
is determined by importance of the work as perceived by other disciplines. However, it neglects authors own
perspective on the different fields and sub fields their paper belongs to. 

PACS numbers in papers contain information on how authors perceive their paper to be belonging to different fields. On PACS hierachical tree,  it is easy to define a distance metric between nodes specified at the same level in the hierarchy from the root. The Weitzman's diversity index \cite{weitzman1992diversity, d1998overview, page2010diversity} is used to characterize degree of 
dissimilarities between the elements of a set. Weitzman diversity ${\cal D} (S)$, measure can be used whenever a clear metric distance is defined between elements of a set $S$. It  is defined as the sum of distances from each element to its nearest neighbor as below: 
\begin{definition}{{\bf Weitzman Diversity} \cite{weitzman1992diversity} }

 Let $\mathbb{U}$ denote the set of PACS codes, and $S, X\subseteq \mathbb{U}$. Let $S=\{u_i, i=1,2,\dots, N\}$ and  $S_k=\{u_i, i=1,2,\dots, k \}$ with $k\leq N$, the distance metric $d(u,v)$ between two elements of a set. The distance between element $u$ and set $X$
is defined as $\bar{d}:  u \times X \rightarrow \mathbb{R}$ such that  
$\displaystyle \bar{d}(u,X)=min_{v\in X} \{d(u,v)\}$ . The Weitzman  diversity ${\cal D }$ of set $S$ is defined as:
 \[ 
 {\cal D } (S)=\sum_{i=1}^{N}\bar{d}(u_i,S_{i-1}) 
 \]
\end{definition}
The $\bar{d}(u_i,S_{k})$ measures the increase in diversity of $S_k$
after the addition of one element $u_i$ \cite{d1998overview}. 
The algorithm to find Weitzman Diversity  \cite{page2010diversity} is as below:
\begin{definition}
 The {\emph Weitzman diversity}, $ {\cal D }(S)$ of a set of elements (or types) given a distance function $\bar{d}(u,X)$ is constructed recursively as follows:
 \begin{enumerate}
  \item[\bf 1:]Let $X=\emptyset$ and initialize $ {\cal D }(X)=0$.
  \item[\bf 2:]Randomly choose an element, $u \in S\setminus X, $, to add to $X$.
  \item[\bf 3:]Find the distance between $u$ and its closet neighbor according to distance $d$.
 i.e, $\bar{d}(u,S)=min_{v \in S}d(u,v)$. Increase $ {\cal D } (X)$ by the $\bar{d}(u,X)$ and add $u$ to the set $X$.
 \item[\bf 4:] If $X \neq S$ go to {\bf 2}.
\end{enumerate}
\end{definition}

\begin{figure}[ht!]
  \centering
  \includegraphics[trim=5.1cm 16.6cm 4.4cm 4.3cm, clip=true, scale=0.7]{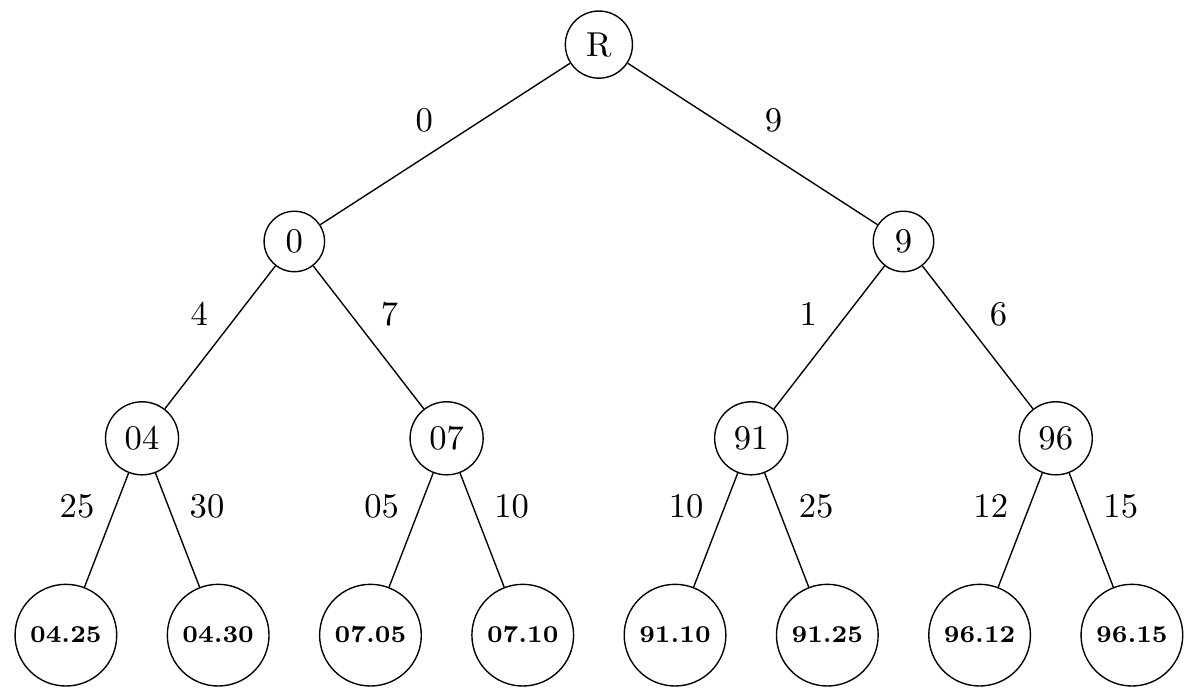}
\caption{Subtree of PACS hierarchical tree}
\label{fig:PACStree} 
\end{figure}

Consider the following illustration on PACS tree (defined up to 3 level) in Fig. \ref{fig:PACStree}. 
Consider the PACS set $S=\{a,b,c\}$, where $a=04.25,b=07.05, c=04.30$ . Initialize $S_0=\{a\}$, ${\cal D }(S_0)=0$ .  Let $S_1=\{a,b\}$,   
${\cal D }(S_1)={\cal D }(S_0)+\bar{d}(b, S_0)=0+d(a,b)=2$, as we need to move two steps backwards to reach a common ancestor (i.e. 0). 
Then ${\cal D }(S)={\cal D }(S_1)+\bar{d}(c, S_1)={\cal D }(S_1)+min \{d(a,c),d(b,c)\}=2+1=3$. Hence ${\cal D }(S)=3$.  

\subsection{Diversity of Papers and Authors}

In this section we compute the diversity of papers published between 1985-2012 in APS journals and their authors using PACS codes.  Diversity of a paper is the  Weitzman diversity of a set $S$, where $S$ is the collection  of PACS codes (paper PACS) mentioned in that paper. 

\begin{figure}[ht!]
  \centering
  \includegraphics[trim=2cm 0.6cm 3.3cm 2.2cm, clip=true, scale=0.35]{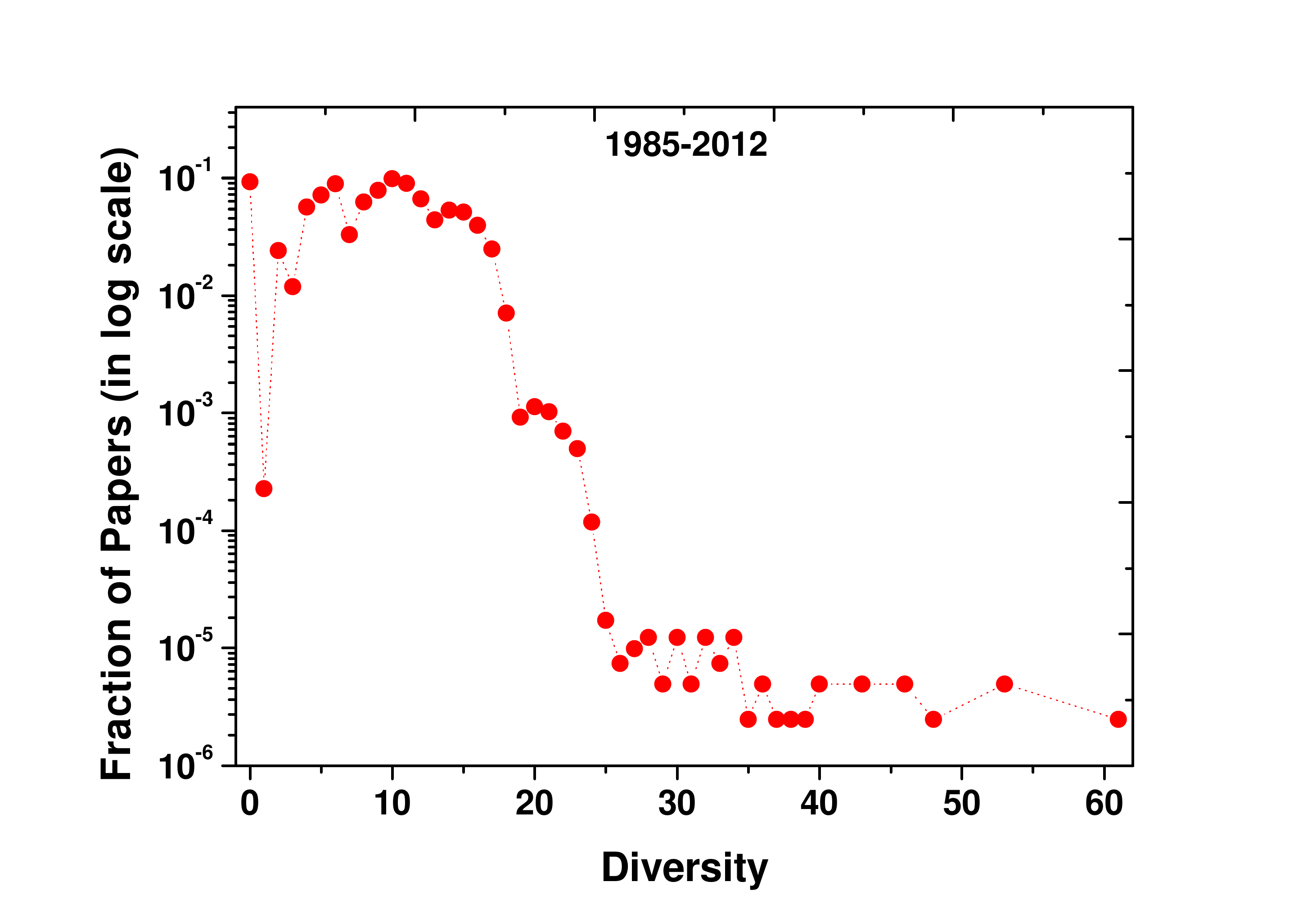}
\caption{Fraction of papers versus diversity}
\label{fig:Diversity Fraction of papers} 
\end{figure}

In Fig \ref{fig:Diversity Fraction of papers}, we show the diversity distribution of papers on a log-linear scale. 
Fraction of papers with ${\cal D}>15$ rapidly declines, and for ${\cal D}\geq 25$, it fluctuates  around $10^{-6}$ to $10^{-5}$. 

\begin{figure}[ht!]
  \centering
  \includegraphics[trim=2cm 0.5cm 3.3cm 2.2cm, clip=true, scale=0.35]{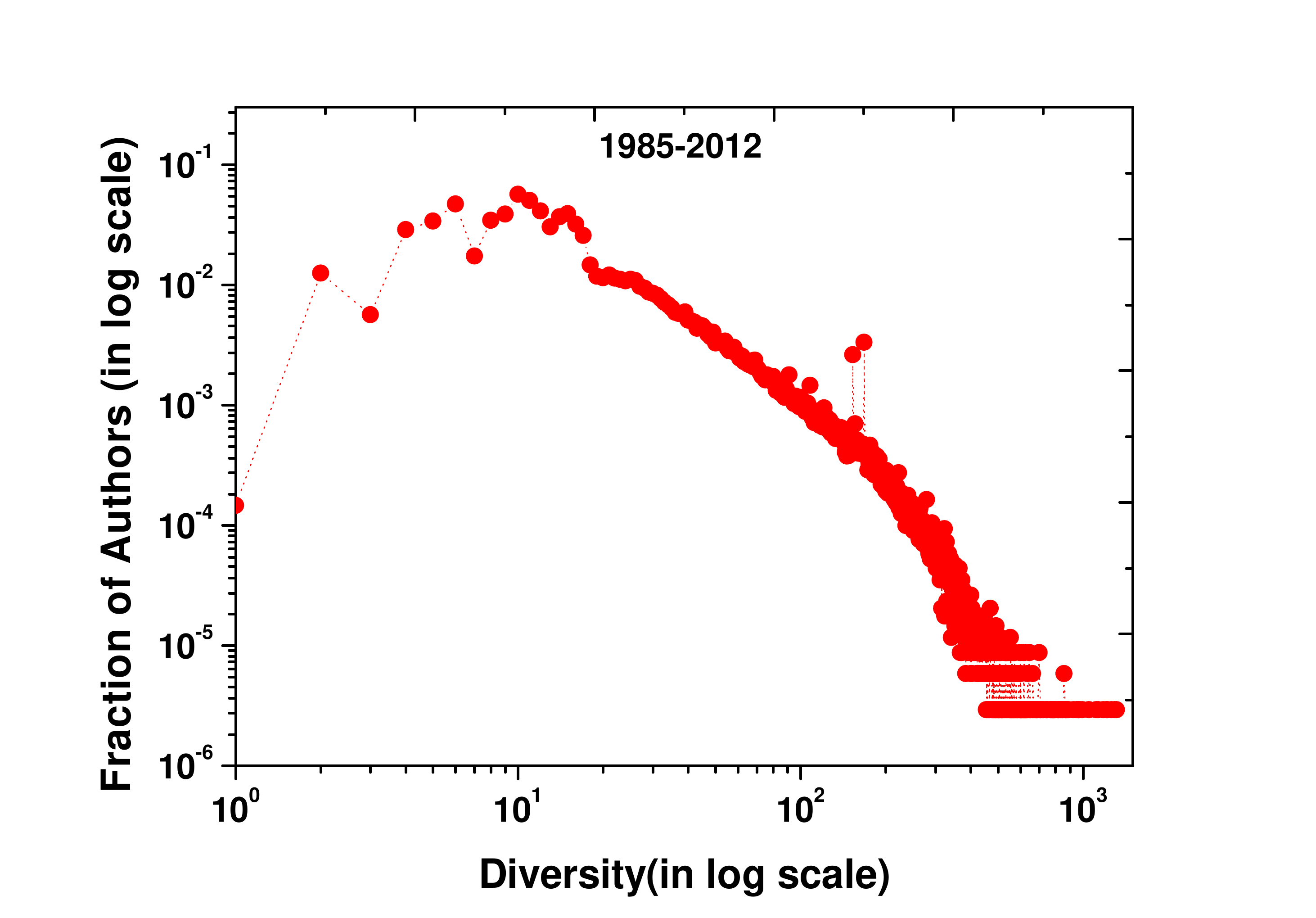}
\caption{Fraction of authors versus diversity}
\label{fig:Diversity Fraction of authors} 
\end{figure}

To calculate the diversity of an author $A$, we take union of all PACS codes $S_{A}$ of papers published by $A$ during a specified time period, and compute the diversity ${\cal D} (S_A)$.  In Fig. \ref{fig:Diversity Fraction of authors}, Weitzman
diversity of authors from 1985-2012 is plotted on a log-log scale. We observe that the fraction of authors with diversity less than ten fluctuates. From diversity ten to about 100 we observe a  power law, and for ${\cal D}>100$, the decay is more rapid. A detailed statistical investigation is yet to be undertaken.
\begin{figure}[ht!]
\centering
  \includegraphics[scale=0.35]{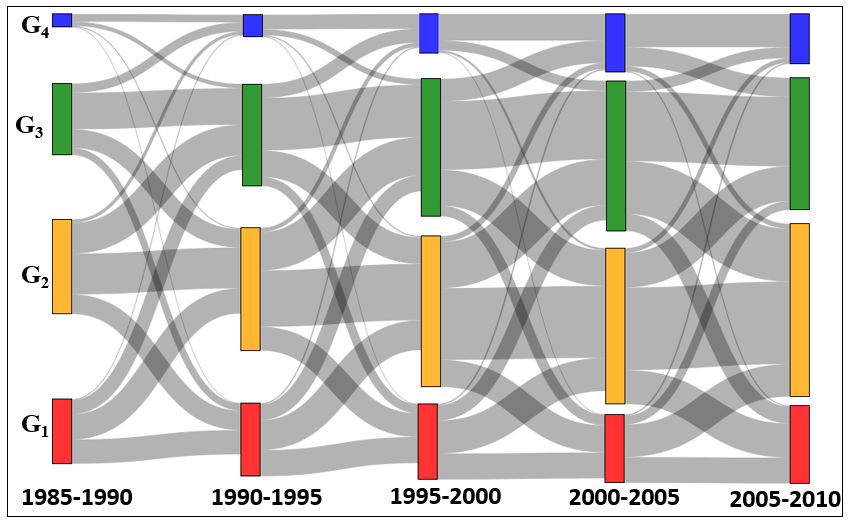}
\caption{Evolution of authors diversity and flows across diversity groups ($G_1=[0,3]$,$G_2=[4,9]$,$G_3=[10,27]$,$G_4=[28+]$) for every 5 years.}
\label{fig:AuthorsDiversity_Alluvial} 
\end{figure}

Unlike the diversity of papers, authors diversity changes overtime as they publish more papers. In Fig.\ref{fig:AuthorsDiversity_Alluvial}, 
we show the time evolution (from 1985-2010) of diversity of authors and their transition from one diversity level to another using alluvial diagram. 
We have binned the authors based on their diversity index $G_i=\{  [0,3], [4,9], [10,27], [28+ ) \}, i=1,2,3,4$. The first and last intervals representing 
lowest ($G_1$, in red) and the highest diversity levels ($G_4$, in blue) and middle levels $G_2$  and $G_3$ are in yellow and green respectively.  
The size of a block indicates the fraction of authors present in that group.  The width of shaded flows corresponds to the fraction of authors 
moving from one group to another. When the width of a group is larger than the incoming flow, the gap indicates the new authors joined in the 
community. We observe that fraction of high diversity authors are increasing over time, where as the proportion 
of low diversity authors is decreasing with time.  Most of the authors have switched to intermediate diversity, including the new authors.

Table \ref{table2}, we show the fraction of authors in different diversity groups from 1985 to 2010. We observe that over time there is a 
steady decrease in fraction of low diversity authors and  increase in high diversity authors, an indication of trend towards interdisciplinary research. 
The movement of high diversity authors is less compared to low diversity authors. 

\begin{table}[ht!]
\renewcommand{\arraystretch}{1.3}
\caption{Fraction of Authors in different groups}
\label{table2}
\centering
\begin{tabular}{|c||c|c|c|c|}
\hline
     & \multicolumn{4}{|c|}{Diversity Groups} \\\hline
Year & $G_1$    & $G_2$   & $G_3$  & $G_4$   \\ \hline
1985-90 & 0.30 & 0.37 & 0.29 & 0.03 \\ \hline
1990-95 & 0.27 & 0.39 & 0.29 & 0.05 \\ \hline
1995-00 & 0.23 & 0.38 & 0.32 & 0.06 \\ \hline
2000-05 & 0.19 & 0.37 & 0.34 & 0.09 \\ \hline
2005-10 & 0.16 & 0.36 & 0.35 & 0.13 \\ \hline
\end{tabular}
\end{table}


\section{Citations and Diversity}
Recently several studies have focused on the statistical characterization of temporal variations of citation received by papers 
\cite{guns2009real, hirsch2007does, radicchi2011rescaling}.  Wang \emph{et al.} have shown that the age at which a paper receives maximum citations follow a 
log-normal distribution \cite{wang2013quantifying}.  In this section we analyze the citations of papers over time and their correlation with diversity. In our data, citations are limited to only those cited by Physical Review(PR)  journals. The actual number of citations of papers may be higher, but we assume that there is substantial correlation between PR and non PR citations.
\begin{figure}[ht!]
  \centering
  \includegraphics[trim=2cm 0.6cm 3.3cm 2.1cm, clip=true, scale=0.35]{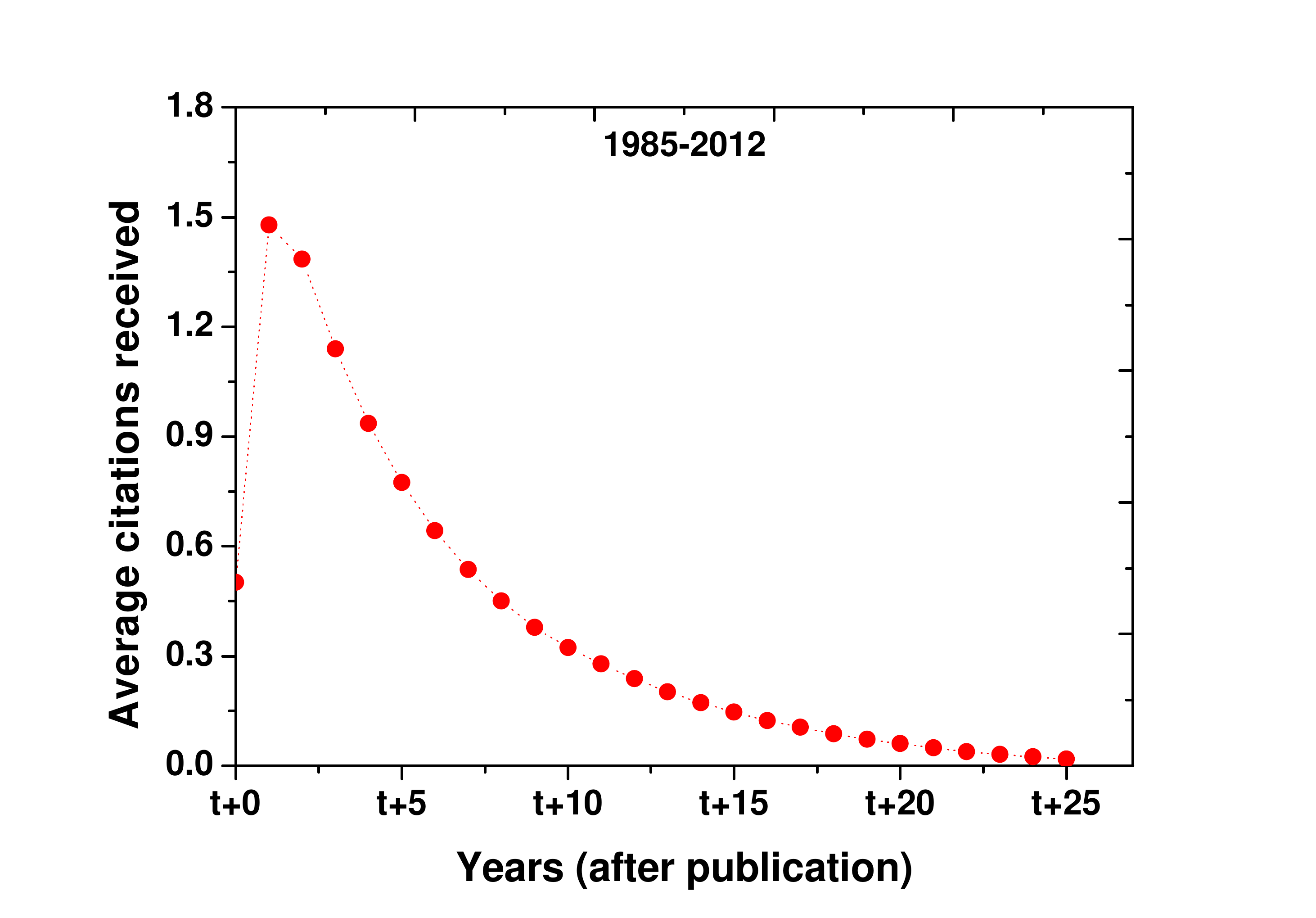}
\caption{Average number of citations received by a paper at the $t^{th}$ year of its publication..}
\label{fig:Averagecitations} 
\end{figure}

\begin{table}[]
\centering
\caption{Percentage of papers for different diversities }
\label{DistributionFit}
\begin{tabular}{|c|c|c|}
\hline
 Year& 1985-1994 & 1994-2003  \\ \hline
Diversity   & papers ($\%$)     & papers ($\%$)                \\ \hline
0         & 13.9                   & 9.7                    \\ \hline
1         & 10.2                     & 9.0                  \\ \hline
2         & 15.5                     & 13.6                       \\ \hline
3         & 19.8                     & 19.0                       \\ \hline
4         & 14.6                     & 16.3                      \\ \hline
5         & 12.5                     & 14.5                       \\ \hline
6         & 8.0                      & 9.8                       \\ \hline
7         & 3.3                      & 4.7                       \\ \hline
8         & 1.7                      & 2.5                        \\ \hline
9 and above  & 0.5                   & 0.9                      \\ \hline
\end{tabular}
\end{table}
\begin{figure*}[ht!]
\centering
 \includegraphics[trim=2.0cm 17.5cm 0.9cm 2.5cm, clip=true, scale=1.0]{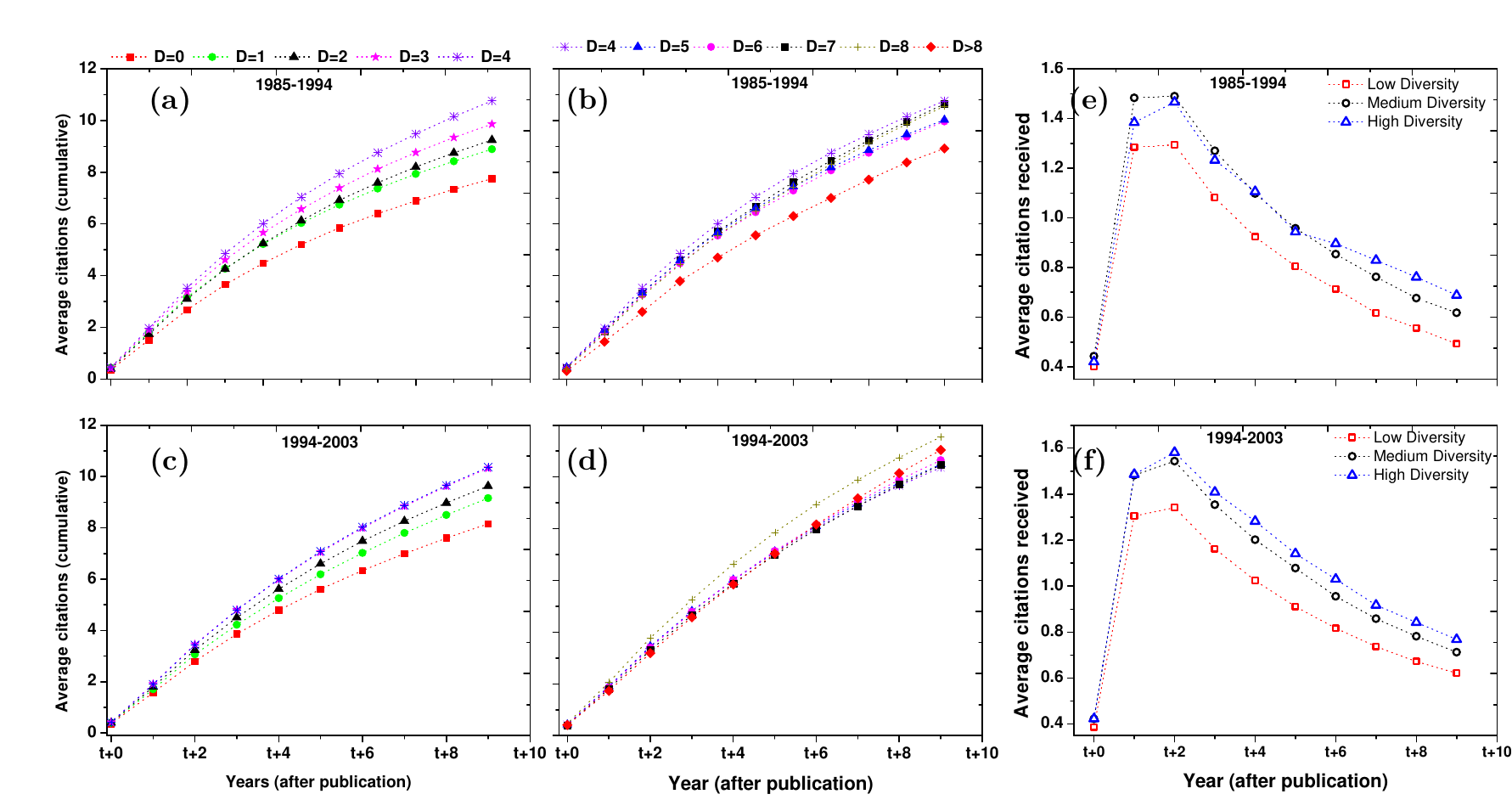}
  \caption{Diversity and Citations: Panels a, b, c and d, shows the cumulative average citations for articles of various diversities 
  for the time periods 1985-1994 and 1994-2003. Diversity groupings [0,4] as in \{a,c\} and [4,8+] as in \{b, d\} have been separated 
  for capturing the trend clearly. Panels (e) and (f) show average number of citations received by a paper from publication year, 
  with diversities grouped into three categories: low($D=\{0,1,2\}$), medium ($D=\{3,4,5\}$) and high ($D=\{6,7,8,8+\}$).   }
  \label{fig:Fig_Divvccumulative}
\end{figure*}
\begin{figure*}[ht!]
\centering
 \includegraphics[trim=2.0cm 20cm 1cm 2.5cm, clip=true, scale=0.9]{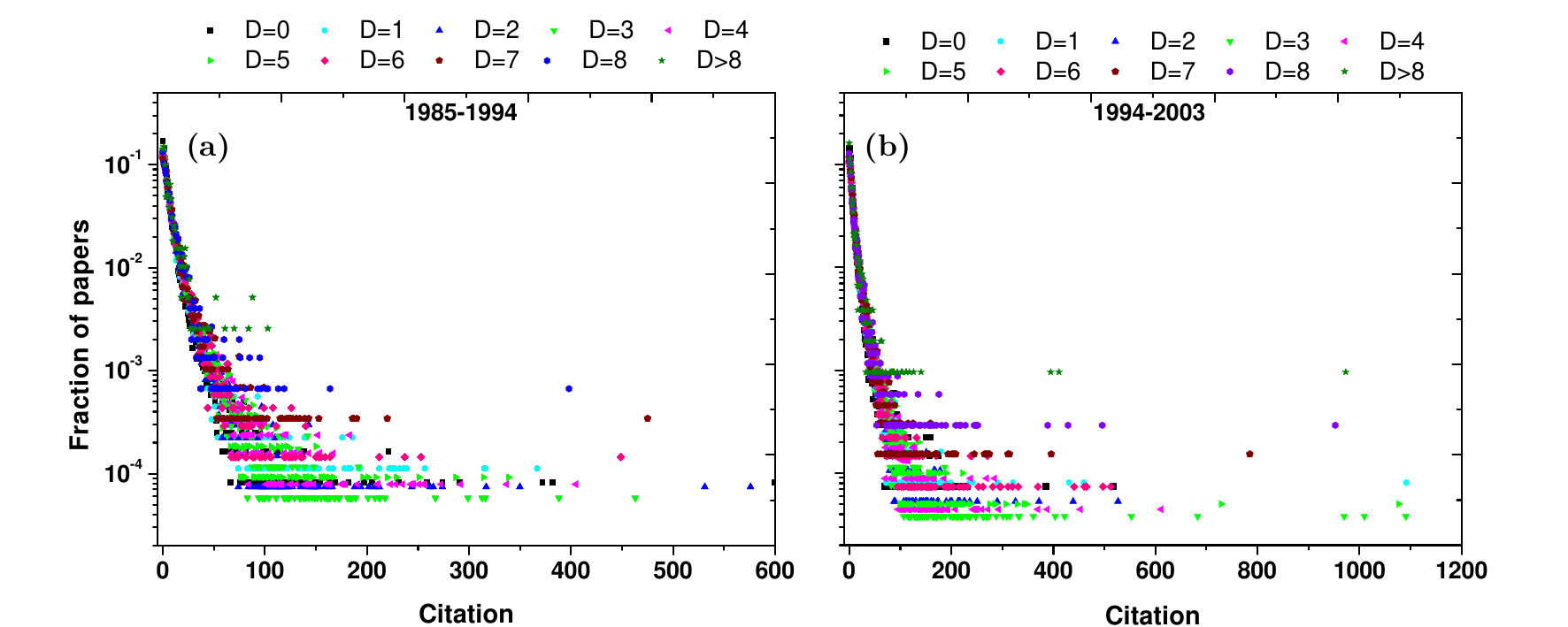}
  \caption{Fraction of papers versus their citations for different diversities. (a) 1985-1994 (b) 1994-2003.}
  \label{fig:fractionofpapers}
\end{figure*}

In Fig. \ref{fig:Averagecitations}, we plot the average number of citations received by papers published in PR journals between 1985-2012. 
It takes about a year for papers to be discovered and cited, there after the number of citations per year continually decreases. 
In Table \ref{DistributionFit}, we show the percentage of papers corresponding to each diversity index between 1985-1994 and 1994-2003. 
We see that in both the cases, maximum number of papers have diversity three. 

Based on this data we try to investigate whether the diversity of papers have strong influence on their citations.  We have grouped papers published 
between 1985-1994 and 1994-2003 according to their diversity index and analyzed their citation pattern for each diversity group for ten years from their 
publication time. In  Fig. \ref{fig:Fig_Divvccumulative} panels (a) and (c), cumulative average citations for diversities zero to four is shown. We notice that citations monotonically increase with the diversity across ten years from their date of publication. In panels (b) and (d), we plot the same 
for diversities four to eight and above. In Panel (b), for years 1985-1994, the cumulative average number of citations received increases till diversity 4 and then begins to decline. It shows that \emph{papers which are too diverse are likely to gather fewer citations}. Such papers may be difficult to follow by focused research groups, and are likely to have lesser depth and relevance to a specific discipline. For papers published between 1994-2003 (Panel c), the citations monotonically increase with the diversity from zero to four. For diversity four to seven (Panel d), the cumulative increase in average citations is roughly the same. Maximum citations is received for diversity eight, from then on wards it declines. This may be partly due to average shift towards higher diversity in 1994-2003 compared to 1985-1994. 

To capture the aggregate  trends of diversity and citations, we further grouped the diversity index into three bins $\{[0,1,2],[3,4,5], [6,7,8,8+]\}$ denoting low, medium and high diversity respectively. The average citations received by these groups of papers from their published year is plotted in Fig. \ref{fig:Fig_Divvccumulative}, panels (e) and (f).  
We see that for 1985-1994, medium diversity papers receive on an average higher citations than the high diversity in the initial years (0-5 yrs) from 
their publication time, where as high diversity papers receive more citations in later years (5 and above). This  indicates that more diverse papers take time to gather citations, in agreement with a recent \emph{Nature} special issue report \cite{Richard2015inter}.  For the years, 1994-2003, if we use the same bins for diversity grouping, the trend seems to be different. Higher diversity groups receive on an average higher citations in all years from their publication year. However, this trend may require careful analysis with a clear definition of groupings.

The analysis so far, depended on the average citations received for papers at a given diversity. However, this does not capture the effects of diversity on citation distribution. Generally large fraction of papers have zero or low citations, and citation distribution is unimodal and single tailed.  We study the citation distribution for each 
diversity in Fig. \ref{fig:fractionofpapers} panels (a) and (b), for time periods 1985-1994 and 1994-2003.  Almost 95\% of the papers 
have citations below 50 in their first 10 years after their publication. Up to 25 citations, distributions follow  exponential decay and later 
the fall is not as steep. This feature is similar across different diversities. A detailed statistical investigation is needed to 
understand the type of distribution and its parameters for different diversities.

\section{Summary and Future Directions}
In this work we have used Weitzman diversity measure to study the diversity profile of scientific articles and authors by using hierarchical structure of PACS codes in APS journals. We have studied the evolution of diversity of authors using alluvial diagram and observed that there is significant monotonic increase in high diversity authors from 1985 to 2010.  The main purpose of our current work is to understand whether being more diverse means having more impact on scientific literature or not? To address this, we studied the correlation between the diversity of papers and their citations. We find that in general high diversity papers receives more citations, but too much diversity can reduce their total citations. We also find that among papers published between 1985-1994, higher diversity paper take longer time to gather citations than the medium and low diversity papers. However, this trends was not found to be universal and depended on the time period and binning of diversity. 

Our work was restricted to only Physical Review journal articles which had hierarchical subject classification through PACS codes. Also, the diversity measure we used are based on the existence of distance metric on the subject classifications. It would be interesting to 
see whether these conclusions are valid even for other data sets such as DBLP, and when we use other diversity measures. 


\section{Acknowledgment}

  The authors would like to thank Anirban Dasgupta for insightful comments and suggestions, and American Physical Society for providing us with the meta data of all Physical Review journal articles and their citations. Enduri would like to thank fellowship and financial grant received from Tata Consultancy Services (TCS) for this work.



%
%
%

\bibliographystyle{IEEEtran}
\bibliography{refs}

\begin{thebibliography}{10}
\providecommand{\url}[1]{#1}
\csname url@samestyle\endcsname
\providecommand{\newblock}{\relax}
\providecommand{\bibinfo}[2]{#2}
\providecommand{\BIBentrySTDinterwordspacing}{\spaceskip=0pt\relax}
\providecommand{\BIBentryALTinterwordstretchfactor}{4}
\providecommand{\BIBentryALTinterwordspacing}{\spaceskip=\fontdimen2\font plus
\BIBentryALTinterwordstretchfactor\fontdimen3\font minus
  \fontdimen4\font\relax}
\providecommand{\BIBforeignlanguage}[2]{{%
\expandafter\ifx\csname l@#1\endcsname\relax
\typeout{** WARNING: IEEEtran.bst: No hyphenation pattern has been}%
\typeout{** loaded for the language `#1'. Using the pattern for}%
\typeout{** the default language instead.}%
\else
\language=\csname l@#1\endcsname
\fi
#2}}
\providecommand{\BIBdecl}{\relax}
\BIBdecl

\bibitem{stehr2000practising}
N.~Stehr and P.~Weingart, \emph{Practising interdisciplinarity}.\hskip 1em plus
  0.5em minus 0.4em\relax University of Toronto Press, 2000.

\bibitem{bordons2005analysis}
M.~Bordons, F.~Morillo, and I.~G{\'o}mez, ``Analysis of cross-disciplinary
  research through bibliometric tools,'' in \emph{Handbook of quantitative
  science and technology research}.\hskip 1em plus 0.5em minus 0.4em\relax
  Springer, 2005, pp. 437--456.

\bibitem{zitt2005facing}
M.~Zitt, ``Facing diversity of science: A challenge for bibliometric
  indicators,'' \emph{Measurement: Interdisciplinary Research and
  Perspectives}, vol.~3, no.~1, pp. 38--49, 2005.

\bibitem{van2002assessment}
A.~F. Van~Raan and T.~N. Van~Leeuwen, ``Assessment of the scientific basis of
  interdisciplinary, applied research: application of bibliometric methods in
  nutrition and food research,'' \emph{Research Policy}, vol.~31, no.~4, pp.
  611--632, 2002.

\bibitem{porter2006interdisciplinary}
A.~L. Porter, J.~D. Roessner, A.~S. Cohen, and M.~Perreault,
  ``Interdisciplinary research: meaning, metrics and nurture,'' \emph{Research
  evaluation}, vol.~15, no.~3, pp. 187--195, 2006.

\bibitem{huutoniemi2010analyzing}
K.~Huutoniemi, J.~T. Klein, H.~Bruun, and J.~Hukkinen, ``Analyzing
  interdisciplinarity: Typology and indicators,'' \emph{Research Policy},
  vol.~39, no.~1, pp. 79--88, 2010.

\bibitem{wagner2011approaches}
C.~S. Wagner, J.~D. Roessner, K.~Bobb, J.~T. Klein, K.~W. Boyack, J.~Keyton,
  I.~Rafols, and K.~B{\"o}rner, ``Approaches to understanding and measuring
  interdisciplinary scientific research (idr): A review of the literature,''
  \emph{Journal of Informetrics}, vol.~5, no.~1, pp. 14--26, 2011.

\bibitem{leydesdorff2011indicators}
L.~Leydesdorff and I.~Rafols, ``Indicators of the interdisciplinarity of
  journals: Diversity, centrality, and citations,'' \emph{Journal of
  Informetrics}, vol.~5, no.~1, pp. 87--100, 2011.

\bibitem{shi2011diversity}
Q.~Shi, B.~Xu, X.~Xu, Y.~Xiao, W.~Wang, and H.~Wang, ``Diversity of social ties
  in scientific collaboration networks,'' \emph{Physica A: Statistical
  Mechanics and its Applications}, vol. 390, no.~23, pp. 4627--4635, 2011.

\bibitem{rafols2010diversity}
I.~Rafols and M.~Meyer, ``Diversity and network coherence as indicators of
  interdisciplinarity: case studies in bionanoscience,'' \emph{Scientometrics},
  vol.~82, no.~2, pp. 263--287, 2010.

\bibitem{Richard2015inter}
V.~N. Richard, ``Interdisciplinary research by the numbers,'' \emph{Nature
  Magazine}, vol. 525, pp. 306--307, 2015.

\bibitem{chakraborty2015understanding}
T.~Chakraborty, V.~Tammana, N.~Ganguly, and A.~Mukherjee, ``Understanding and
  modeling diverse scientific careers of researchers,'' \emph{Journal of
  Informetrics}, vol.~9, no.~1, pp. 69--78, 2015.

\bibitem{schmidt2006methodological}
M.~Schmidt, J.~Glaser, F.~Havemann, and M.~Heinz, ``A methodological study for
  measuring the diversity of science,'' 2006.

\bibitem{liu2010mining}
L.~Liu, F.~Zhu, C.~Chen, X.~Yan, J.~Han, S.~Y. Philip, and S.~Yang, ``Mining
  diversity on networks,'' in \emph{Database Systems for Advanced
  Applications}.\hskip 1em plus 0.5em minus 0.4em\relax Springer, 2010, pp.
  384--398.

\bibitem{pan2012evolution}
R.~K. Pan, S.~Sinha, K.~Kaski, and J.~Saram{\"a}ki, ``The evolution of
  interdisciplinarity in physics research,'' \emph{Scientific reports}, vol.~2,
  2012.

\bibitem{chakraborty2013automatic}
T.~Chakraborty, S.~Kumar, M.~D. Reddy, N.~Ganguly, and A.~Mukherjee,
  ``Automatic classification and analysis of interdisciplinary fields in
  computer sciences,'' in \emph{Social Computing (SocialCom), 2013
  International Conference on}.\hskip 1em plus 0.5em minus 0.4em\relax IEEE,
  2013, pp. 180--187.

\bibitem{martin2013coauthorship}
T.~Martin, B.~Ball, B.~Karrer, and M.~E.~J. Newman, ``Coauthorship and citation
  patterns in the physical review,'' \emph{Phys. Rev. E}, vol.~88, p. 012814,
  Jul 2013.

\bibitem{redner2005citation}
S.~REDNER, ``Citation statistics from 110 years of physical review,''
  \emph{Physics today}, vol.~58, no.~6, pp. 49--54, 2005.

\bibitem{page2010diversity}
S.~E. Page, \emph{Diversity and complexity}.\hskip 1em plus 0.5em minus
  0.4em\relax Princeton University Press, 2010.

\bibitem{reed2004double}
W.~J. Reed and M.~Jorgensen, ``The double pareto-lognormal distribution—a new
  parametric model for size distributions,'' \emph{Communications in
  Statistics-Theory and Methods}, vol.~33, no.~8, pp. 1733--1753, 2004.

\bibitem{weitzman1992diversity}
M.~L. Weitzman, ``On diversity,'' \emph{The Quarterly Journal of Economics},
  pp. 363--405, 1992.

\bibitem{d1998overview}
C.~T. d'Arnoldi, J.-L. Foulley, and L.~Ollivier, ``An overview of the weitzman
  approach to diversity,'' \emph{Genetics Selection Evolution}, vol.~30, no.~2,
  pp. 149--161, 1998.

\bibitem{guns2009real}
R.~Guns and R.~Rousseau, ``Real and rational variants of the h-index and the
  g-index,'' \emph{Journal of Informetrics}, vol.~3, no.~1, pp. 64--71, 2009.

\bibitem{hirsch2007does}
J.~E. Hirsch, ``Does the h index have predictive power?'' \emph{Proceedings of
  the National Academy of Sciences}, vol. 104, no.~49, pp. 19\,193--19\,198,
  2007.

\bibitem{radicchi2011rescaling}
F.~Radicchi and C.~Castellano, ``Rescaling citations of publications in
  physics,'' \emph{Physical Review E}, vol.~83, no.~4, p. 046116, 2011.

\bibitem{wang2013quantifying}
D.~Wang, C.~Song, and A.-L. Barab{\'a}si, ``Quantifying long-term scientific
  impact,'' \emph{Science}, vol. 342, no. 6154, pp. 127--132, 2013.

\end{thebibliography}
%

\end{document}